\documentclass[showpacs,aps]{revtex4}
\usepackage{amssymb}
\usepackage{amsmath}
\usepackage{graphicx}
\usepackage{lscape}
\usepackage{booktabs}
\usepackage{epsfig}
\begin{document}
\title{Charged particle elliptic flow in p+p collisions at LHC energies \\in
        a transport model PACIAE}
\author{Dai-Mei Zhou$^{1}$, Yu-Liang Yan$^{2}$,Bao-Guo Dong$^{2}$, Xiao-Mei
Li$^{2}$, Du-Juan Wang$^{1}$, Xu Cai$^{1}$ and Ben-Hao Sa$^{1,2,3}$ }
\address{1 Institute of Particle Physics and Key Laboratory of Quark $\&$ Lepton Physics,
 Huazhong Normal University, Wuhan 430079, China\\
2 China Institute of Atomic Energy, P.O. Box 275
(18), Beijing 102413, China \\
3 CCAST (World Laboratory), P. O. Box 8730 Beijing 100080, China}

\begin{abstract}
The parton and hadron cascade model PACIAE based on PYTHIA was used to
investigate the charged particle elliptic flow in minimum bias pp collisions
at the LHC energies. The strings were distributed randomly in the transverse
ellipsoid of the pp collision system with major axis of $R$ (proton radius)
and minor axis of $R(1-\xi)$ before parton rescattering. The charged particle
elliptic flow as a function of the random number $\xi$ and transverse
momentum $p_T$ were investigated. The calculated $v_2/\varepsilon$ as a
function of reaction energy increases monotonously with increasing reaction
energy up to $\sqrt{s}\sim$7 TeV and then turns to saturation. With the
parton-parton cross section enlarges three times in parton rescattering, the
rapidity integrated charged particle elliptic flow may reach 0.025 at $p_T
\sim$2 GeV/c in the minimum bias pp collisions at $\sqrt{s}$=7 TeV.
\end{abstract}

\maketitle

Elliptic flow is the 2nd harmonic in fourier expansion of the particle
momentum azimuthal distribution \cite{posk}
\begin{equation}
\frac{1}{2\pi}\frac{dN}{dp_Tdyd\phi}\\
=\frac{1}{2\pi}\frac{dN}{p_Tdp_Tdy}[1+2\sum_nv_ncosn(\phi-\phi_R)].
\end{equation}
In theoretical studies the elliptic flow is calculated by
\begin{equation}
v_2=\langle\overline{[\frac{p_x^2-p_y^2}{p_T^2}]}\rangle,
\end{equation}
where $\overline{O}$ indicates an average of operator $O$ over particles in
an event and $\langle O\rangle$ an average over events \cite{sa2}.

The elliptic flow was suggested in hydrodynamic calculations \cite{JYOll1}
as a signature of collective expansion in the relativistic nuclear collisions.
The consistency between experimental data of $v_1(y)$ as well as $v_2(p_T)$
at mid-rapidity and the corresponding hydrodynamic predictions was regarded
as an evidence of the production of partonic matter in relativistic
nucleus-nucleus collisions \cite{miclo,csr}. The elliptic flow is now one of
the most important observables in relativistic nucleus-nucleus collisions
\cite{star,phenix,phobos,alice1}. It has attracted high attention both
experimentally and theoretically.

In non-central collision the reaction zone (overlap region) between two
colliding nuclei is spatially asymmetric. The rescattering process among
the produced particles transfers this spatial asymmetry into the momentum
space, and the momentum distribution of the produced particles becomes
azimuthally anisotropic. This picture is generally valid for a large system.
In the pp collision at RHIC energy, the average multiplicity $\langle n_{ch}
\rangle$ is not big enough for collective effects to be detected. However, in
the pp collisions at $\sqrt{s}$=7 TeV, it has already been observed that the
multiplicity can reach $\frac{dN_{ch}}{d\eta}>30$ \cite{alice,cms}. Therefore
the elliptic flow may be measurable in the pp collision at LHC energies and
the further theoretical studies are also required.

The fluctuations in pp interaction region can result in a sizable spatial
eccentricity was assumed in \cite{wiedemann}. With further assumptions on the
nature of this fluctuations and on the eccentricity scaling of elliptic flow,
they reported that the elliptic flow becomes measurable in high-multiplicity
pp collisions at LHC energies. The initial interactions produce a number of
hot spots was assumed in \cite{chaudhuri}. Their hydrodynamical
evolution of two or more hot spots generated a sufficiently large
elliptic flow. In \cite{avsar} a Monte Carlo implementation of the
dipole (gluon) cascade model was introduced to study the elliptic
flow in pp collision at LHC energies. They predicted that $v_2$ in
$\sqrt{s}$= 7 TeV pp collision is around 6-7\%. The ideal 3+1D
hydrodynamic simulation was used to study the $v_2$ in
\cite{ortona}, and they concluded that the elliptic flow can occur
at least for top multiplicities pp collision at $\sqrt{s}$= 14 TeV.
In \cite{bozek} the hydrodynamic model calculations showed that if
high multiplicity events in the pp collisions at LHC energies contain two
flux tubes, the elliptic flow may be observable. Many other models also gave
their predictions \cite{cunq}.

We studied the charged particle elliptic flow in minimum bias pp collisions
at the LHC energies by a parton and hadron cascade model PACIAE \cite{sa}
based on PYTHIA \cite{soj2} in this letter. It is well known that PYTHIA
is a model for high energy hadron-hadron (hh) collisions. In the PYTHIA model
a hh collision is decomposed into the parton-parton collisions. A hard
parton-parton collision is described by the lowest leading order perturbative
QCD (LO-pQCD). The soft parton-parton collision is considered empirically.
Because the initial- and final-state QCD radiations and multiparton
interactions are considered in the parton-parton scattering, the consequence
of a hh collision is a parton multijet configuration composed of di-quarks
(anti-diquarks), quarks (anti-quarks), and gluons, besides a few hadronic
remnants. This parton multijet configuration is followed by the string
construction and fragmentation (hadronization). Therefore one obtains a
hadronic final state for a hh (pp) collision.

For pp collisions the PACIAE model is different from PYTHIA in the addition
of the parton rescattering before hadronization and the hadron rescattering
after hadronization. The PACIAE model consists of the parton initialization,
parton evolution (rescattering), hadronization, and hadron evolution
(rescattering) four stages.

\begin{enumerate}

\item The parton initialization: \\
The parton initialization is performed by the PYTHIA model with
string fragmentation (hadronization) switched-off. One obtains a
parton configuration composed of quarks, anti-quarks, and gluons,
besides a few hadronic remnants for a pp (hh) collision after
diquarks (anti-diquarks) being split randomly into quarks
(anti-quarks). This parton configuration is regarded as
quark-gluon matter (QGM) formed in the initial state (fireball) of
pp collision.

\item The parton evolution (rescattering): \\
The rescattering among partons in QGM is then considered by the 2
$\rightarrow$ 2 LO-pQCD parton-parton cross sections \cite{comb}. The
differential cross section of a subprocess $ij\rightarrow kl$ is
\begin{equation}
\frac{d\sigma_{ij\rightarrow
kl}}{d\hat{t}}=K\frac{\pi\alpha_s^2}{\hat{s}}\sum_{ij\rightarrow
kl},
\end{equation}
where the factor $K$ (assumed to be 3 in this letter) is introduced
considering the higher order and the nonperturbative corrections,
$\alpha_s$= 0.47 stands for the effective strong coupling constant, and
$\hat{s}$, $\hat{t}$, as well as $\hat{u}$ refer to the Mandelstam
variables. The specific subprocess $q_1\bar q_2\rightarrow q_1\bar q_2$,
for instance, has
\begin{equation}
\sum_{q_1\bar q_2\rightarrow q_1\bar q_2}
=\frac{4}{9}\frac{\hat{s}^2+\hat{u}^2}{\hat{t}^2}.
\label{eq3}
\end{equation}
It is regularized by introducing the parton color screen mass
$\mu$ (=0.63 GeV) as
\begin{equation}
\sum_{q_1\bar q_2\rightarrow q_1\bar q_2}
=\frac{4}{9}\frac{\hat{s}^2+\hat{u}^2}{(\hat{t}-\mu^2)^2}.
\end{equation}
The total cross section of $i+j$ parton collision is then
\begin{equation}
\sigma_{ij}(\hat{s})=\sum_{k,l}\int_{-\hat{s}}^{0}d\hat{t}\;
\frac{d\sigma_{ij\to kl}}{d\hat{t}}.
\end{equation}
With the total and differential cross sections above the parton
evolution (rescattering) can be simulated by the Monte Carlo
method until all parton-parton collisions are exhausted (partonic
freeze-out). The results introduced later were calculated using
this pQCD cross sections except special mention.
\item The hadronization: \\
In the hadronization stage, the partonic matter (QGM) formed after
parton rescattering is hadronized by the Lund string fragmentation
regime \cite{soj2} after string reconstruction or by the Monte
Carlo coalescence model. The former model has been described in
detail in \cite{soj2}. We proposed a phenomenological coalescence
model for the later one \cite{sa}. In this model, two partons
coalesce a meson and three partons a baryon (antibaryon) according
to the valence quark structure of hadron and the flavors,
positions as well as the momenta of coalescing partons. The
positions and momenta of coalescing partons are constrained to the
phase space requirement
  \begin{equation}
  \frac{16\pi^2}{9}\Delta r^3\Delta p^3=\frac{h^3}{d}
  \end{equation}
where $h^3/d$ is the volume occupied by a single hadron in the
phase space , $d$=4 refers to the spin and parity degeneracies of
the hadron, $\Delta r$ and $\Delta p$ stand for the position and
momentum distances between coalescing partons, respectively. In
addition, the momenta of coalescing partons have to satisfy the
momentum conservation.
\item The hadron evolution (rescattering): \\
In this stage the hadronic matter after hadronization proceeds
rescattering. It is dealt with by the usual two-body elastic and
inelastic collisions \cite{sa1}, until the hh collision pairs are
exhausted (hadronic freeze-out). The rescatterings among $\pi, K,
p, n, \rho (\omega), \Delta, \Lambda, \Sigma , \Xi, \Omega,
J/\Psi$ and their antiparticles are considered for the moment. The
isospin averaged parametrization formula \cite{koch,bald} is
assumed for cross section of $hh$ collisions. In addition, an
assumed constant total cross sections ($\sigma_{\rm{tot}}^{NN}=40$~mb,
$\sigma_{\rm{tot}}^{\pi N}=25$~mb, $\sigma_{\rm{tot}}^{KN}=20$~mb, and
$\sigma_{\rm{tot}}^{\pi \pi}=10$ ~mb) and ratio of inelastic to
total cross section (0.85) are provided as another option.
\end{enumerate}

\vspace{0.5in}
\begin{figure}[ht]
\begin{center}
\epsfig{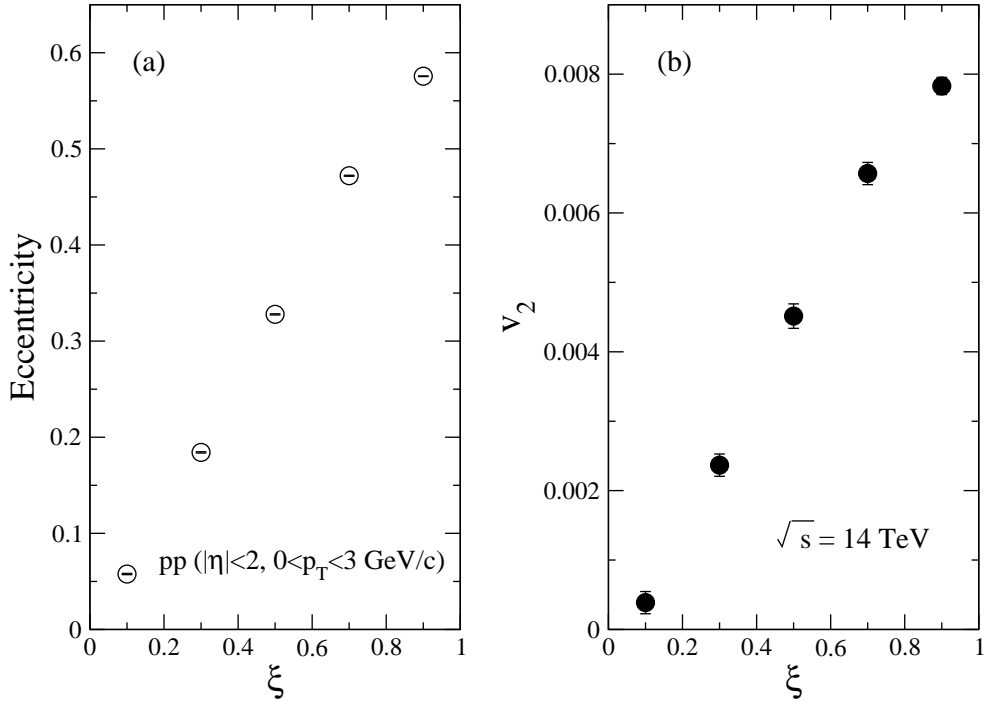}
\end{center}
\vspace{-0.1in} \caption{Charged particle eccentricity and elliptic flow as
a function of $\xi$ in minimum bias pp collision at $\sqrt{s}$ =14 TeV.}
\label{v2_xi}
\end{figure}

The PYTHIA and PACIAE models with default parameters (except the factor $K$
was assumed to be 3) were used to generate pp minimum bias events at
$\sqrt{s}$=14 TeV. We found that the charged particle $v_2(p_T)$ from the
PYTHIA calculations was almost zero and the one from PACIAE was just around
0.0012. Following the idea of hot spots \cite{wiedemann,chaudhuri,bozek} we
assumed that the strings before parton rescattering were distributed randomly
in the transverse elliptic zone of the pp collision system. The major and
minor axes of this ellipsoid were
\begin{equation}
a=R,
\end{equation}
and
\begin{equation}
b=R(1-\xi),
\end{equation}
respectively. In above equations $R$ was the radium of proton and the
parameter $\xi$ (random number) reflects the fluctuation in the initial
position distribution in pp collision. The constituent partons in a string
were sampled randomly in a circle centered at the string and with radius of
1 fm.

We calculated the charged particle eccentricity \cite{phob1}
\begin{equation}
\langle\varepsilon\rangle=\langle\frac{\sigma^{2}_{y}-\sigma^2_x}{\sigma^2_y+
\sigma^2_x}\rangle
\end{equation}
according to parton initial state in position space (later it was simply
indicated as $\varepsilon$). In this equation $\sigma^2_x=\overline{x^2}-
\overline{x}^2$ and $\sigma^2_y =\overline{y^2}-\overline{y}^2$ were the
variance of the initial parton distribution in the $x$ and $y$ directions in
a given event.

\begin{figure}[ht]
\begin{center}
\epsfig{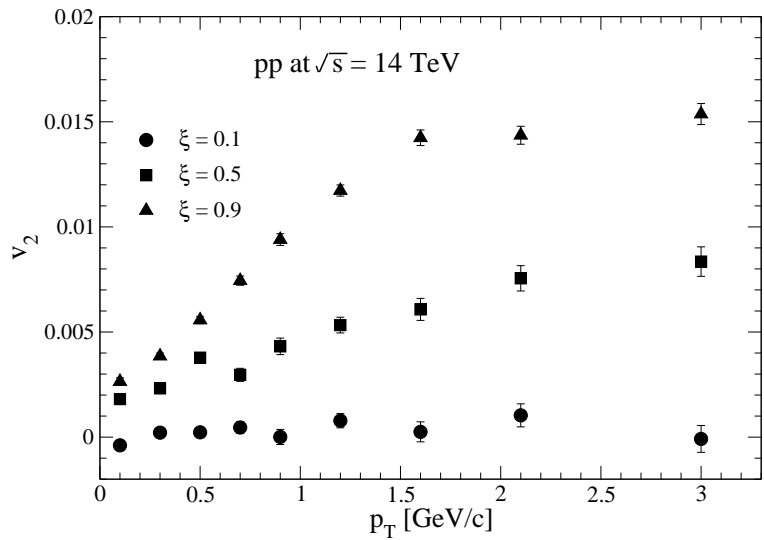}
\end{center}
\vspace{-0.1in} \caption{Charged particle $v_2$ as a function of $p_T$ in
minimum bias pp collisions at $\sqrt{s}$=14 TeV.}
\label{v2_pt_xi}
\end{figure}
Figure \ref{v2_xi} (a) and (b) give the charged particle
eccentricity and the integrated elliptic flow $v_2$ ($|\eta|<2$
and $0<p_T<3 GeV/c$) as a function of $\xi$ in the minimum bias pp
collisions at $\sqrt{s}$=14 TeV in the PACIAE calculations,
respectively. Both the eccentricity and elliptic flow increase
with increasing $\xi$ that may mean the integrated $v_2$ is
proportional to the initial eccentricity as expected in
\cite{Olli}.

The charged particle integrated elliptic flow $v_2$ as a function of the
transverse momentum $p_T$ in minimum bias pp collisions at $\sqrt{s}$=14 TeV
is given in Fig.\ref{v2_pt_xi}. In this figure the full circles, squares, and
triangles are the results calculated with $\xi$=0.1, 0.5, and 0.9,
respectively. The $v_2$ value increases with increasing $\xi$ that may
represent the final momentum asymmetry is originated from initial position
anisotropy.

\begin{figure}[ht]
\begin{center}
\epsfig{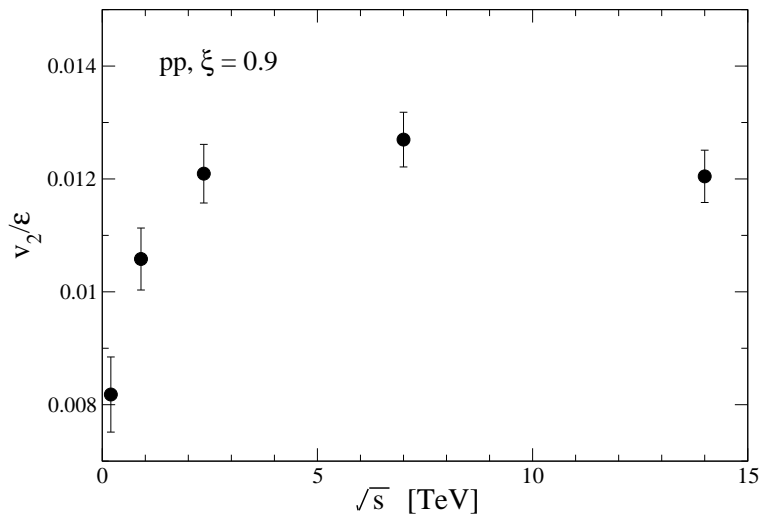}
\end{center}
\vspace{-0.1in} \caption{The scaled charged particle elliptic flow,
$v_2/ \varepsilon$, as a function of collision energy in minimum
bias pp collisions .} \label{v2ecc}
\end{figure}

Figure \ref{v2ecc} shows the reaction energy dependence of scaled
elliptic flow $v_2/\epsilon$ in the minimum bias pp collisions
calculated with $\xi=0.9$. In this figure $v_2$ is integrated over
$\eta$ and $p_T$. One sees here that $v_2/\varepsilon$ increases
with increasing reaction energy monotonously up to 7 TeV and then
turns to saturation. This behavior is similar to $v_2/\epsilon$
varies with centrality observed in relativistic heavy ion
collisions reported in \cite{dres}.

\vspace{0.5in}
\vspace{0.3in}
\begin{figure}[ht]
\begin{center}
\epsfig{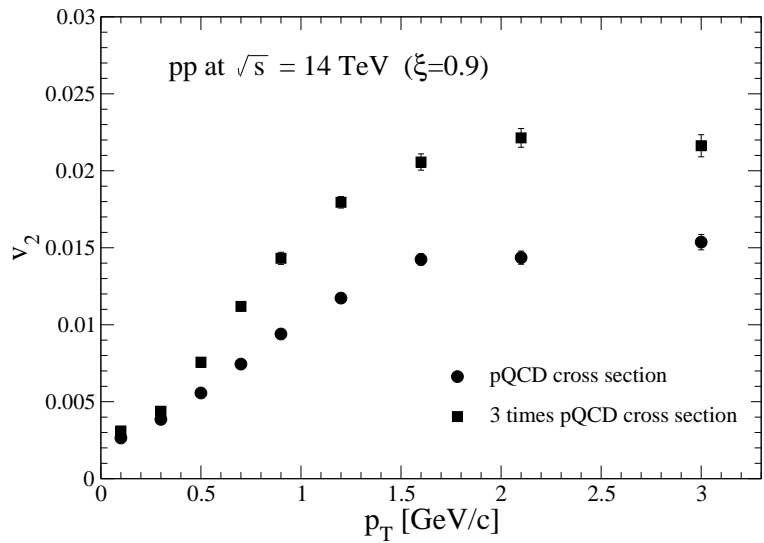}
\end{center}
\vspace{-0.1in} \caption{Charged particle elliptic flow $v_2$ as a function
of the transverse momentum $p_T$ in minimum bias pp collision at $\sqrt{s}$=
14 TeV with two different parton-parton cross sections.}
\label{v2p}
\end{figure}
The function of charged particle integrated elliptic flow $v_2$
versus transverse momentum $p_T$ varying with parton-parton
interaction cross section in the minimum bias pp collision at
$\sqrt{s}$=14 TeV calculated with $\xi$=0.9 is shown in Fig.
\ref{v2p}. The full circles are the results calculated by the pQCD
parton-parton cross sections in the parton rescattering while the
full squares are calculated by three times enlarged parton-parton
cross sections. We see in this figure that the charged particle
integrated elliptic flow increases nearly 60\% at $p_T\sim$2 GeV/c
if the parton-parton cross section is increased by three times.
That indicates the importance of the parton rescattering.

In summary, the parton and hadron cascade model PACIAE was used to study the
charged particle elliptic flow in the minimum bias pp collisions at LHC
energies. The strings were distributed randomly in the transverse ellipsoid
of the pp collision system with major axis of $R$ (proton radius) and minor
axis of $R(1-\xi)$ before parton rescattering. We investigated the charged
particle elliptic flow as a function of the random number $\xi$ and
transverse momentum $p_T$, respectively. The calculated $v_2/\varepsilon$ as
a function of reaction energy first increases monotonously with increasing
reaction energy up to $\sqrt{s}\sim$7 TeV and then turns to saturation. We
find that the parton rescattering is also very important. If the
parton-parton cross section enlarges three times, the rapidity integrated
charged particle elliptic flow $v_2$ increases nearly 60\% and reach to
0.025 at $p_T\sim$2 GeV/c in the minimum bias pp collisions at $\sqrt{s}$=7
TeV.

\acknowledgments

The financial supports from NSFC (10975062,11075217, 11047142,
10705012) and from the Commission on Higher Education in China are
acknowledged.


\end{document}